\documentclass[conference]{IEEEtran}
\usepackage{amsmath,graphicx,hyperref,multirow}
\usepackage{fancyhdr} 
\usepackage{amssymb}
\usepackage{booktabs}
\title{Self-Improvement for Audio Large Language Model using Unlabeled Speech}

\author{
  Shaowen Wang*, Xinyuan Chen*, Yao Xu* \\
  University of Illinois at Urbana-Champaign \\
  \texttt{shaowen2.uillinois@gmail.com; xinyuan.c@yahoo.com; yaoxu5.uillinois@gmail.com}
}

\begin{document}

\maketitle
\renewcommand{\thefootnote}{\fnsymbol{footnote}}
\footnotetext[1]{*These authors contributed equally.}

\vspace{-0.8em}
{\small\textit{This paper has been accepted to Interspeech 2025. This is a preprint version.}}

\begin{abstract}

Recent audio LLMs have emerged rapidly, demonstrating strong generalization across various speech tasks. However, given the inherent complexity of speech signals, these models inevitably suffer from performance degradation in specific target domains. To address this, we focus on enhancing audio LLMs in target domains \textbf{without} any labeled data. We propose a self-improvement method called SI-SDA, leveraging the information embedded in large-model decoding to evaluate the quality of generated pseudo labels and then perform domain adaptation based on reinforcement learning optimization. Experimental results show that our method consistently and significantly improves audio LLM performance, outperforming existing baselines in WER and BLEU across multiple public datasets of automatic speech recognition (ASR), spoken question-answering (SQA), and speech-to-text translation (S2TT). Furthermore, our approach exhibits high data efficiency, underscoring its potential for real-world deployment.
\end{abstract}

\section{Introduction}

In recent years, large language models (LLMs) have emerged as general-purpose task solvers in NLP community~\cite{brown2020language}, and also have been playing an increasingly vital role in human life. Meanwhile, the Transformer-based architecture has proven remarkable performance for processing human speech~\cite{gulati2020conformer} with high sampling rate. Consequently, enabling a unified model to handle acoustic and textual inputs has attracted surge of interest in the both speech and multimodal communities, which has given rise to notable advances in audio LLMs~\cite{chu2023qwen,hu2024wavllm,tang2023salmonn,zhang2024speechgpt}. \par
However, due to the inherent variability of speech signals, audio LLMs often experience performance degradation in specific acoustic environments, such as ASR tasks under unseen background noise~\cite{fathullah2024prompting}. Existing solutions typically rely on collecting domain-specific speech data, annotating it, and then fine-tuning the model with supervision. Alternatively, some methods adopt data augmentation to simulate data from the target domain~\cite{chen2022noise}. Inspired by the self-improvement capabilities exhibited by LLMs~\cite{huang2022large}, this work explores a fundamental and challenge question: Can audio LLMs achieve self-improvement without relying on target domain annotations? In other words, given only a pre-trained audio LLMs and a few hours or less of unlabeled speech from the target domain, is it possible for this model to acquire the robustness of target domain in a self-training~\cite{lee2023feature} manner. \par
To address this question, we direct our focus on reinforcement learning (RL), which has been demonstrated its effectiveness on sequence generation optimization task in recent research~\cite{rennie2017self}. Typically, RL optimization is viewed as a post-training scheme widely used in LLMs, a.k.a., reinforcement learning with human feedback (RLHF)~\cite{gu2024mutual}. Instead of relying on human feedback for annotation, we argue that the autoregressive model inherently conveys information about the quality of its generated samples during decoding, particularly when hallucinations occur~\cite{huang2023opera}. \par
Inspired by this, we propose a self-improvement speech domain adaptation (SI-SDA) to improve the performance of audio LLMs in multiple tasks and domains. By analyzing the information flow in audio LLM, we find that the attention matrix of deep Transformer layer is inherent quality indicator for samples generated by auto-regressive decoding, and SI-SDA leverage this information to perform group policy optimization in a beam set. Subsequently, the model undergoes back propagation by maximizing the log-probability of the those chosen sequences, thereby improving its robustness to the target domain. Notably, the entire optimization process of SI-SDA requires no additional annotated data and does \textbf{not} rely on recalling source domain data, as is commonly done in traditional unsupervised domain adaptation (UDA) methods~\cite{ganin2015unsupervised,sun2017unsupervised}. Instead, model determines which self-generated samples should be encouraged or suppressed through the uncertainty during its auto-regressive decoding, which is consistent with human's learning process~\cite{hu2024self}. \par
To validate the effectiveness of SI-SDA, we conduct extensive experiments on three key tasks in audio LLMs: automatic speech recognition (ASR), spoken question-answering (SQA), and speech-to-text translation (S2TT) across multiple datasets. In the ASR task, our method requires only a few hours of data to enhance audio LLM performance in challenging conditions such as background noise and speaker accents, achieving up to an 18\% relative WER reduction. Similarly, in ST and SQA, SI-SDA improves generation quality in terms of BLEU scores in a efficient manner. Furthermore, the promising performance of SI-SDA validates the self-improvement ability of audio LLMs, and pave the way for its practical application in the real world.

\section{Related Work}
Inspired by the success of LLMs, researchers have explored their potential to enhance speech-to-text tasks. Early studies focused on single tasks such as ASR~\cite{fathullah2024prompting}, AST~\cite{huang2023speech}, and spoken language understanding~\cite{zhu2024zero}, where an audio encoder extracts semantic representations for LLM processing. Driven by the availability of large-scale data, more audio-specific LLMs have been developed~\cite{das2024speechverse,rubenstein2023audiopalm,du2023lauragpt}. These models can be categorized based on speech representation methods: Discrete token-based models: An audio codec discretizes speech, expanding the LLM’s vocabulary to enable direct audio-text interaction~\cite{zhang2023speechgpt,nguyen2025spirit}. Continuous representation-based models: A pre-trained encoder, such as an ASR model encoder or a self-supervised learning model, processes raw waveforms. The extracted features are then combined with LLM word embeddings via a modality adapter, allowing seamless integration of audio and text~\cite{bapna2022mslam, chu2024qwen2}.
\section{Methodology}
\subsection{Salience Score in Audio LLM}
To extract acoustic representations from time domain waveform, mainstream audio LLMs usually employ a pre-trained speech encoder to obtain a audio sequence $X = \{x_1, x_2, \cdots, x_3 \}$ in a Transformer-like feature $X \in \mathbb{R}^{L \times D}$, where L is the length and D is dimension that needs to be aligned with LLM's dimension. Then X is fed into a decoder-only Transformer for text sequence $Y$ generation. Typically, the next token prediction training pattern is utilized as shown in follows:   
\begin{equation}
    P(Y \mid X) = \prod_{t=1}^{T} P(y_t \mid Y_{1:t-1}, X; \theta). 
\label{eq1}
\end{equation}
The LLM is then trained to maximize the likelihood of each predicted token \( y_t \), conditioned on both the preceding text tokens \( y_{<t} \) and the audio sequence \( X \), parameterized by \( \theta \). \par
Despite down-sampling on audio sequence, $X$ can be still longer than $Y$ due to high sampling rate of speech signal. Furthermore, the alignment between $X$ and $Y$ is complex: in human language, a short syllable can sometimes correspond to multiple text tokens, leading to an imbalance in token-to-token mapping. In this work, we utilize the saliency score to analyze the contribution of $X$ and $Y_{0:t-1}$ when predicting $y_t$ in time step $t$. Following common practice, we use the Taylor expansion~\cite{michel2019sixteen} to calculate the saliency score~\cite{simonyan2013deep} for each element of the attention matrix:
\begin{equation}
    I_l = \left| \sum_h A_{h,l} \odot \frac{\partial \mathcal{L}(x)}{\partial A_{h,l}} \right|.
    \label{eq:saliency}
\end{equation}
where \( A_{h,l} \) represents the attention matrix of the \( h \)-th attention head in the \( l \)-th layer, and \( x \) denotes the input. The function \( \mathcal{L}(x) \) is the loss function for the given task, such as the cross-entropy objective in classification problems. The saliency matrix \( I_l \) is computed by summing over all attention heads and taking the element-wise product with the gradient of the loss with respect to the attention weights. The absolute value ensures that the computed importance scores remain positive. The matrix \( I_l(i,j) \) quantifies the significance of information flow from the \( j \)-th token to the \( i \)-th token in the speech-to-text task. \par
\subsection{Token-level Distribution Analysis with Saliency Score}
\label{sec:saliency_correct_wrong}
After defining the saliency score $I_l$ in Eq.~(\ref{eq:saliency}), we first investigate whether there are systematic differences in the saliency distribution when the model decodes a token correctly versus incorrectly, where ASR task is taken for analysis due to the availability of its unique ground truth. 
Let $\mathcal{C}$ be the set of time steps at which the model outputs the correct token, and $\mathcal{E}$ be the set of time steps at which the model makes an error.
We compute the saliency matrix $I_l(i,j)$ for each decoder layer $l$, where $i$ corresponds to the \emph{query} position (i.e., the current token being predicted), and $j$ indexes the \emph{key} positions in either the audio sequence $X$ or the previously generated tokens $Y_{<t}$ (including prompt tokens).
To quantify how strongly a particular token position $i$ relies on the special prompt tokens, we sum up the saliency values over those prompt positions (such as ``\verb|<transcribe>|"):
\begin{equation}
    S_{\mathrm{prompt}}(i) \;=\; 
    \sum_{j \,\in\, \mathrm{Prompt}} \; 
    I_l(i,j),
\end{equation}
and normalize by the total saliency over all possible $j$:
\begin{equation}
    R(i) = \frac{S_{\mathrm{prompt}}(i)}{\sum_{j} I_l(i,j)}.
    \label{eq:R_prompt}
\end{equation}
Here, $R(i)$ serves as an approximate measure of ``prompt dependence'' at the $i$-th decoding step. 
We then average $R(i)$ over $\mathcal{C}$ and $\mathcal{E}$:
\begin{align}
    \overline{R}_{\mathcal{C}} &= \frac{1}{|\mathcal{C}|}\sum_{i \in \mathcal{C}} R(i), \\
    \overline{R}_{\mathcal{E}} &= \frac{1}{|\mathcal{E}|}\sum_{i \in \mathcal{E}} R(i).  
\end{align}
In practice, we observe that $\overline{R}_{\mathcal{E}}  \overline{R}_{\mathcal{C}}$ (see Section 5.1), this indicates that \emph{erroneous} tokens generally rely more on the prompt region in terms of saliency, implying that the model might be ``falling back'' to prompt tokens rather than aligning with the appropriate acoustic frames or local context.

\subsection{Saliency-Based Indicator for Hypothesis Quality}
\label{sec:saliency_indicator}

While the analysis above relies on comparing correct versus incorrect tokens, in many real-world scenarios we only have hypotheses $\hat{Y}$ without ground-truth references.
We propose an unsupervised indicator that leverages saliency distributions to gauge the overall quality of a hypothesis at the sequence level.

Let $\hat{Y} = \{\hat{y}_1, \hat{y}_2, \dots, \hat{y}_T\}$ be any candidate decoded from $X$ by the model.
We compute the saliency matrix $I_l(i,j)$ for each token position $i \in \{1, 2, \dots, T\}$, then define a ``prompt reliance'' function:
\begin{equation}
    R(i) =\frac{\sum_{j \,\in\, \mathrm{Prompt}} I_l(i,j)}{\sum_{j} I_l(i,j)},
\end{equation}
as in Eq.~(\ref{eq:R_prompt}), but now applied to a \emph{single} hypothesis where we do not know which tokens are correct or incorrect.
To capture the \emph{overall} dependence on prompt tokens, we aggregate $R(i)$ across all positions in $\hat{Y}$:
\begin{equation}
    Q(\hat{Y}) = \frac{1}{T}\sum_{i=1}^{T} R(i),
    \label{eq:Qy}
\end{equation}
which yields a normalized scalar indicating how much, on average, the model's saliency is tied to the prompt.
Intuitively, if the decoding is strongly influenced by the prompt rather than aligning with the local audio frames, $Q(\hat{Y})$ will be larger, suggesting a potentially lower-quality hypothesis.

In a batch of $N$-best hypotheses $\{\hat{Y}^1, \hat{Y}^2, \dots, \hat{Y}^N\}$ generated by beam search algorithm, we can rank these candidates by their $Q(\hat{Y}^n)$ values or apply a threshold $\tau$ to reject those with abnormally high $Q(\hat{Y})$.
Since $Q(\hat{Y})$ is purely computed from the model's internal saliency maps, it requires no external ground-truth labels and acts as an \emph{unsupervised} measure of alignment self-consistency.
In later sections, we demonstrate that this simple metric correlates with actual decoding errors and can serve as a helpful indicator for RL-based optimization. \par
\subsection{Reward Design and Optimization}
\label{sec:rl_optimization}

After obtaining $Q(\hat{Y})$ as an unsupervised measure of hypothesis quality in Eq.~(\ref{eq:Qy}), we can further treat it as a \emph{reward} signal in a reinforcement learning (RL) setting, reminiscent of minimum WER training~\cite{prabhavalkar2018minimum}. Let $\{\hat{Y}^1, \ldots, \hat{Y}^N\}$ be the $N$ candidate hypotheses generated from beam search or sampling for a given input $X$. 
We interpret the model's decoding process as a \emph{policy} $\pi_\theta(\hat{Y}\mid X)$ parameterized by $\theta$, 
where $\pi_\theta(\hat{Y}^n\mid X) = P(\hat{Y}^n\mid X; \theta)$.
We then define the reward for each hypothesis $\hat{Y}^n$ to be its saliency-based quality score $Q(\hat{Y}^n)$.

\noindent \textbf{Baseline and Advantage}. To stabilize the training and reduce variance, we adopt a baseline $b$, commonly taken as the average reward over the $N$ hypotheses, and then define the \emph{advantage} function $A^n$ for each hypothesis $\hat{Y}^n$ as:
\begin{align}
    A^n = - (Q\bigl(\hat{Y}^n\bigr) - \bar{Q}), \ \ \text{where} \ \bar{Q} = \frac{1}{N}\,\sum_{n=1}^N Q\bigl(\hat{Y}^n\bigr).
\label{eq:advantage}
\end{align}
so that hypotheses with $Q(\hat{Y}^n) < \bar{Q}$ (above-average reward) receive a positive advantage, while below-average ones receive a negative advantage. \par
Following a standard policy gradient formulation, 
we define the RL-style objective $\mathcal{L}_{\mathrm{rl}}(\theta)$ by weighting the log-probabilities of sampled (or beam-searched) hypotheses with their advantages:
\begin{equation}
    \mathcal{L}_{\mathrm{rl}}(\theta)= - \sum_{n=1}^N 
    \Bigl[
        A^n \;\log P\bigl(\hat{Y}^n \mid X;\theta\bigr)
    \Bigr] 
    \label{eq:rl_loss}
\end{equation}
Intuitively, hypotheses with higher reward (lower $Q(\hat{Y}^n)$) than the baseline are reinforced by increasing their log-probabilities, whereas those with lower reward are suppressed.
Minimizing $\mathcal{L}_{\mathrm{RL}}(\theta)$ encourages the model to shift probability mass towards hypotheses whose saliency-based indicator $Q(\hat{Y})$ suggests a more coherent alignment with the acoustic frames, and away from those that rely excessively on prompt tokens or exhibit other undesirable saliency patterns.

\section{Experiment}
\subsection{Models and Dataset}
We use Qwen-Audio2~\cite{chu2024qwen2} as the representative audio LLM, with ASR as the primary research task. Since Qwen-Audio already achieves strong zero-shot performance on clean datasets (WER on LibriSpeech is 1.6), we follow the STAR setting~\cite{hu2024self} and define the target datasets as those involving background noise, speaker accents, or specific application scenarios. Specifically, we evaluate our approach on CHiME-4~\cite{vincent2016chime4}, LibriSpeech with FreeSound (0dB)~\cite{font2013freesound}, CommonVoice~\cite{ardila2019common}, Switchboard~\cite{2020swb}, and TED3. Meanwhile, we explore the extension on S2TT task on CoVoST2~\cite{wang2020covost}, and SQA task with LibriSQA~\cite{zhao2024librisqa}. Notably, only the training set is employed for sampling and optimization, and test set is only used for evaluation.
\subsection{Training, baselines, and Evaluation} \label{baseline}
Due to computational constraints, we adopt LoRA-tuning with rank = 16 and a learning rate of 1e-5, training for 2 epochs per dataset with Adam optimizer. All experiments are conducted on two NVIDIA A100 (40GB) GPUs. The beam size is set as 10 and keep top 5 after de-duplication. \par
We compare our method against the following baselines:
\begin{itemize}
    \item Zero-shot performance: The audio LLM's performance on the test set with beam search, where no adaptation method is performed.
    \item Self-train: The model is trained using top-1 generated samples in a beam set, the training objective is cross-entropy loss.
    \item Filtering: Before self-training, we discard 20\% of utterances in training set with high diversity, as it typically indicates model has high uncertainty on this data point~\cite{hu2024self}. The diversity is measured by edit distance between each other.
    \item Conf: Based on self-training, but with confidence-based re-weighting applied during cross-entropy computation.
    \item STAR: An attention score-based re-weighting method, combining with utterance filtering. 
    \item SFT: Fine-tuning on labeled target data, typically considered the upper bound for UDA methods. For fair comparison, LoRA tuning is used that is consistent with proposed method. 
\end{itemize}
For the ASR task, we report the performance in terms of WER and WER reduction (WERR). For S2TT and SQA, we use BLEU score to evaluate model performance. 
\section{Results}
\subsection{Main Results on ASR}
We report the main results of SI-SDA and baselines in Table~\ref{table1}, where the best results are in bold. From Table~\ref{table1}, it is observed that (1) our method achieves the best performance on all test sets except for the ``car" subset. Compared to the limited improvements achieved by other UDA baselines, SI-SDA demonstrates larger WER reductions across all test sets, achieving up to 20.8\% WERR on TED-3. (2) Although the zero-shot performance varies significantly, ranging from 3.2 to 25.6, our method consistently achieves notable improvements, demonstrating the stability of SI-SDA across different scenarios. (3) Compared to SFT, SI-SDA converges faster, typically requiring only 2 to 5 hours before performance on the validation set plateaus. This is primarily because self-training samples are generated by the model itself, rather than forcing the model to fit a new data distribution, as in SFT. On one hand, this demonstrates strong data efficiency, only hours of unlabeled speech is required for adaptation. On the other hand, it highlights potential future research directions, particularly on how to further improve performance when large amounts of unlabeled speech data are available. \par

\begin{table*}[t]
\caption{WER Main result on ASR task. The baselines are introduced in Section~\ref{baseline}. The percentage values in the bottom-right corner represent WERR. ``SFT" column utilizes the same amount of \textit{labeled} data and can be considered the upper bound performance.}
\centering
\resizebox{1.88\columnwidth}{!}{
\begin{tabular}{cc|cccccc|c}
\toprule
Test set & Domain & Zero-shot & Self-train & Filtering & Conf  & STAR & SI-SDA & SFT \\  \midrule
CHiME-4  &  real noises  & 6.6 & 6.6 & 6.4&6.4 & 6.0 & $\textbf{5.7}_{-15.8\%}$ & 5.0  \\ \midrule
\multirow{3}{*}{LS-Freesound} & car & 3.2 & 3.5 & 3.5 & 3.2 &\textbf{2.4} & $2.5_{-21.9\%}$ &2.5 \\
 & airport & 11.9 & 11.7 &11.7 & 11.8 & 11.0 & $\textbf{10.2}_{-14.3\%}$ & 7.8 \\ 
 & babble  & 25.6 & 23.9 &24.4 & 23.2 & 22.6 & $\textbf{20.7}_{-19.1\%}$  & 14.7 \\ \midrule
\multirow{4}{*}{CommonVoice} & Singapore & 5.8 & 5.6 & 5.7 & 5.5 & 5.3 & $\textbf{5.0}_{-13.8\%}$ & 3.9 \\
& Africa  & 5.6   & 5.5 & 5.4 & 5.5 & 5.2 & $\textbf{4.9}_{-12.5\%}$ & 4.6  \\
& Australian & 4.9 & 4.7 & 4.7 & 4.6 & 4.3 & $\textbf{4.3}_{-14.3\%}$ & 3.7\\
& Indian  &  6.1 & 6.1 & 6.0 & 5.8 & \textbf{5.5} & $\textbf{5.5}_{-9.8\%}$ & 4.0\\\midrule
SwitchBoard & telephone & 11.1 & 10.9 & 11.8 & 10.7 & 9.9 & $\textbf{9.8}_{-11.7\%}$ & 8.8 \\\midrule
TED-3 & TED talk & 7.2 & 7.0 & 7.1 & 6.6 & 6.0 & $\textbf{5.7}_{-20.8\%}$ & 5.2  \\ 
\bottomrule
\end{tabular}}
\label{table1}
\end{table*}

\begin{figure}[t]
  \begin{center}
  \includegraphics[scale=0.22]{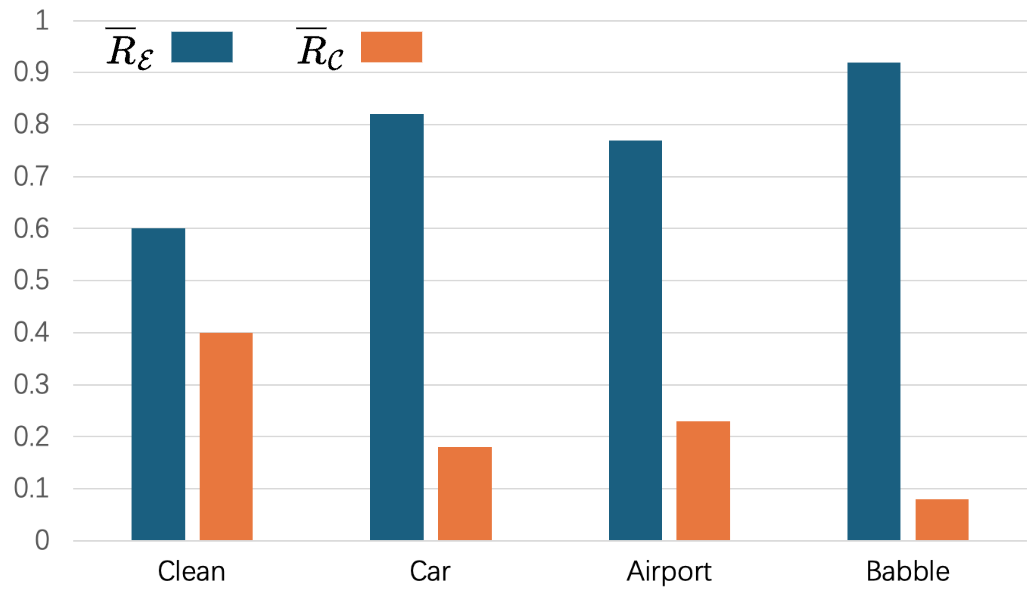}
  \end{center}
  \caption{The $\overline{R}_{\mathcal{E}}$ and $\overline{R}_{\mathcal{C}}$ comparison across clean and noise ASR datasets. All values are normalized such that their sum equals 1, allowing for a clearer observation of the relative proportions. }
  \label{f1}
\end{figure}

To validate the relationship between $\overline{R}_{\mathcal{E}}$ and $\overline{R}_{\mathcal{C}}$, we report the proportion of them on both clean LibriSpeech test set LS-Freesound (mix LibriSpeech with different background noises) for comparison. As visualized in Figure~\ref{f1}, $\overline{R}_{\mathcal{E}}$ are obviously larger than $\overline{R}_{\mathcal{C}}$ across these for test set. This indicates that whenever a token-level issue arises, the model tends to rely more heavily on the prompt instruction rather than the input sequence during decoding. From a deeper perspective, this reflects an adaptive yet suboptimal degradation strategy in audio LLMs when encountering uncertain or low-confidence inputs: the model attempts to ``re-query" the global context or initial state in search of better decoding cues.  

\subsection{Ablation Study on RL}
Given a beam set of hypotheses with different Q-values, this ablation study aims to examine the effectiveness of the adopted RL strategy. To this end, we compare against the following baselines:
(1) Min-Q: Uses the hypothesis with the lowest Q-value in the beam set as a pseudo label for fine-tuning.
(2) Re-Atten: Applies attention score-based re-weighting when computing the cross-entropy loss, which is a key component in STAR. (3) DPO: Selects the hypotheses with the highest and lowest Q-values for preference optimization, aiming to maximize implicit reward proposed in~\cite{rafailov2024direct}.\par
From Table~\ref{table2} we observed that all optimization method achieve performance gain compared with zero-shot result, demonstrating the significance of attention-based information during decoding. Meanwhile, SI-SDA surpasses all other baselines by a large margin across different test sets, which indicate that RL is a well-suited approach in this task, particularly because beam search naturally provides pairwise sampled results, making it an effective foundation for optimization.

\begin{table}[t]
\caption{Ablation study on CHiME-4 and LS-Freesound.}
\centering
\resizebox{0.94\columnwidth}{!}{
\begin{tabular}{c|cccccc}
\toprule
Testset & Min-Q & Re-Atten & DPO & SI-SDA \\ \midrule
CHiME-4   & 6.6 & 6.1 & 6.7 & 5.7\\ \midrule
car       & 3.0  & 2.6 &  3.3 & 2.5 \\
airport   &  11.6 & 11.4 & 11.5 & 10.2\\
babble    & 23.6 & 22.8 & 20.8 & 20.7 \\
\bottomrule
\end{tabular}}
\label{table2}
\end{table}

\subsection{Extension on S2TT and SQA}
In this experiment, we investigate the scalability of SI-SDA speech-to-text translation and spoken question answering tasks. Using the CoVoST2 dataset, we evaluate our method on three distinct language pairs from CoVoST2: English to German (en-de), Chinese to English (zh-en), and Italian to English (it-en). For SQA, we focus on Part-I of LirbriSQA dataset, which involves free-form answer generation. We maintain a beam size of 10, but due to the significant differences between hypotheses, we do not perform hypothesis repetition. Instead, we directly compute the loss using Equation (10). \par
The results are summarized in Table~\ref{table3}. It is observed that SI-SDA outperforms the baselines on all test sets. This indicates that even in S2TT and SQA, where labels are not unique, attention-based scores remain effective for evaluating sample quality under unsupervised conditions. Moreover, this suggests that the behavior of audio LLMs in ASR generalizes well across different speech tasks, demonstrating its robustness and adaptability.

\begin{table}[t]
\caption{Extension on S2TT and SQA. All resutls are reported in terms of BLEU score.}
\centering
\resizebox{0.88\columnwidth}{!}{
\begin{tabular}{c|cccccc}
\toprule
Testset   & Zero-shot & Conf & STAR & SI-SDA \\ \midrule
en-de     & 29.8      &  29.2   &  30.7  & 30.7 \\ 
zh-en     & 24.4      &  24.3   &  30.1  & 32.5 \\
it-en     & 35.7      &  36.0   &  36.1  & 36.1 \\ \midrule
part-I    & 21.7      &  16.7   & 20.0   & 22.2 \\
\bottomrule
\end{tabular}}
\label{table3}
\end{table}

\section{Conclusion}
This work focuses on improving audio LLM performance across different target domains without using any labeled data. Based on an analysis of audio LLM decoding behavior, we propose SI-SDA, a RL-based approach for unsupervised domain adaptation that accordingly optimize the generated samples from the N-best list. Experimental results demonstrate that our method consistently enhances performance across various speech tasks, while its data efficiency highlights its practical value for real-world applications, facilitating the deployment of audio LLMs in real-world scenarios.

\bibliographystyle{IEEEtran}
\bibliography{mybib}

\begin{thebibliography}{10}
\providecommand{\url}[1]{#1}
\csname url@samestyle\endcsname
\providecommand{\newblock}{\relax}
\providecommand{\bibinfo}[2]{#2}
\providecommand{\BIBentrySTDinterwordspacing}{\spaceskip=0pt\relax}
\providecommand{\BIBentryALTinterwordstretchfactor}{4}
\providecommand{\BIBentryALTinterwordspacing}{\spaceskip=\fontdimen2\font plus
\BIBentryALTinterwordstretchfactor\fontdimen3\font minus \fontdimen4\font\relax}
\providecommand{\BIBforeignlanguage}[2]{{%
\expandafter\ifx\csname l@#1\endcsname\relax
\typeout{** WARNING: IEEEtran.bst: No hyphenation pattern has been}%
\typeout{** loaded for the language `#1'. Using the pattern for}%
\typeout{** the default language instead.}%
\else
\language=\csname l@#1\endcsname
\fi
#2}}
\providecommand{\BIBdecl}{\relax}
\BIBdecl

\bibitem{brown2020language}
T.~Brown, B.~Mann, N.~Ryder, M.~Subbiah, J.~D. Kaplan, P.~Dhariwal, A.~Neelakantan, P.~Shyam, G.~Sastry, A.~Askell \emph{et~al.}, ``Language models are few-shot learners,'' \emph{Advances in neural information processing systems}, vol.~33, pp. 1877--1901, 2020.

\bibitem{gulati2020conformer}
A.~Gulati, J.~Qin, C.-C. Chiu, N.~Parmar, Y.~Zhang, J.~Yu, W.~Han, S.~Wang, Z.~Zhang, Y.~Wu \emph{et~al.}, ``Conformer: Convolution-augmented transformer for speech recognition,'' \emph{arXiv preprint arXiv:2005.08100}, 2020.

\bibitem{chu2023qwen}
Y.~Chu, J.~Xu, X.~Zhou, Q.~Yang, S.~Zhang, Z.~Yan, C.~Zhou, and J.~Zhou, ``Qwen-audio: Advancing universal audio understanding via unified large-scale audio-language models,'' \emph{arXiv preprint arXiv:2311.07919}, 2023.

\bibitem{hu2024wavllm}
S.~Hu, L.~Zhou, S.~Liu, S.~Chen, L.~Meng, H.~Hao, J.~Pan, X.~Liu, J.~Li, S.~Sivasankaran \emph{et~al.}, ``Wavllm: Towards robust and adaptive speech large language model,'' \emph{arXiv preprint arXiv:2404.00656}, 2024.

\bibitem{tang2023salmonn}
C.~Tang, W.~Yu, G.~Sun, X.~Chen, T.~Tan, W.~Li, L.~Lu, Z.~Ma, and C.~Zhang, ``Salmonn: Towards generic hearing abilities for large language models,'' \emph{arXiv preprint arXiv:2310.13289}, 2023.

\bibitem{zhang2024speechgpt}
D.~Zhang, X.~Zhang, J.~Zhan, S.~Li, Y.~Zhou, and X.~Qiu, ``Speechgpt-gen: Scaling chain-of-information speech generation,'' \emph{arXiv preprint arXiv:2401.13527}, 2024.

\bibitem{fathullah2024prompting}
Y.~Fathullah, C.~Wu, E.~Lakomkin, J.~Jia, Y.~Shangguan, K.~Li, J.~Guo, W.~Xiong, J.~Mahadeokar, O.~Kalinli \emph{et~al.}, ``Prompting large language models with speech recognition abilities,'' in \emph{ICASSP 2024-2024 IEEE International Conference on Acoustics, Speech and Signal Processing (ICASSP)}.\hskip 1em plus 0.5em minus 0.4em\relax IEEE, 2024, pp. 13\,351--13\,355.

\bibitem{chen2022noise}
C.~Chen, N.~Hou, Y.~Hu, S.~Shirol, and E.~S. Chng, ``Noise-robust speech recognition with 10 minutes unparalleled in-domain data,'' in \emph{ICASSP 2022-2022 IEEE International Conference on Acoustics, Speech and Signal Processing (ICASSP)}.\hskip 1em plus 0.5em minus 0.4em\relax IEEE, 2022, pp. 4298--4302.

\bibitem{huang2022large}
J.~Huang, S.~S. Gu, L.~Hou, Y.~Wu, X.~Wang, H.~Yu, and J.~Han, ``Large language models can self-improve,'' \emph{arXiv preprint arXiv:2210.11610}, 2022.

\bibitem{lee2023feature}
J.~Lee and G.~Lee, ``Feature alignment by uncertainty and self-training for source-free unsupervised domain adaptation,'' \emph{Neural Networks}, vol. 161, pp. 682--692, 2023.

\bibitem{rennie2017self}
S.~J. Rennie, E.~Marcheret, Y.~Mroueh, J.~Ross, and V.~Goel, ``Self-critical sequence training for image captioning,'' in \emph{Proceedings of the IEEE conference on computer vision and pattern recognition}, 2017, pp. 7008--7024.

\bibitem{gu2024mutual}
S.~Gu, ``Mutual enhancement of large language and reinforcement learning models through bi-directional feedback mechanisms: A case study,'' \emph{arXiv preprint arXiv:2401.06603}, 2024.

\bibitem{huang2023opera}
Q.~Huang, X.~Dong, P.~Zhang, B.~Wang, C.~He, J.~Wang, D.~Lin, W.~Zhang, and N.~Yu, ``Opera: Alleviating hallucination in multi-modal large language models via over-trust penalty and retrospection-allocation,'' in \emph{Proc. CVPR}, 2024.

\bibitem{ganin2015unsupervised}
Y.~Ganin and V.~Lempitsky, ``Unsupervised domain adaptation by backpropagation,'' in \emph{International conference on machine learning}.\hskip 1em plus 0.5em minus 0.4em\relax PMLR, 2015, pp. 1180--1189.

\bibitem{sun2017unsupervised}
S.~Sun, B.~Zhang, L.~Xie, and Y.~Zhang, ``An unsupervised deep domain adaptation approach for robust speech recognition,'' \emph{NeuroComputing}, vol. 257, pp. 79--87, 2017.

\bibitem{hu2024self}
Y.~Hu, C.~Chen, C.-H.~H. Yang, C.~Qin, P.-Y. Chen, E.~S. Chng, and C.~Zhang, ``Self-taught recognizer: Toward unsupervised adaptation for speech foundation models,'' \emph{arXiv preprint arXiv:2405.14161}, 2024.

\bibitem{huang2023speech}
Z.~Huang, R.~Ye, T.~Ko, Q.~Dong, S.~Cheng, M.~Wang, and H.~Li, ``Speech translation with large language models: An industrial practice,'' \emph{arXiv preprint arXiv:2312.13585}, 2023.

\bibitem{zhu2024zero}
Z.~Zhu, X.~Cheng, H.~An, Z.~Wang, D.~Chen, and Z.~Huang, ``Zero-shot spoken language understanding via large language models: A preliminary study,'' in \emph{Proceedings of the 2024 Joint International Conference on Computational Linguistics, Language Resources and Evaluation (LREC-COLING 2024)}, 2024, pp. 17\,877--17\,883.

\bibitem{das2024speechverse}
N.~Das, S.~Dingliwal, S.~Ronanki, R.~Paturi, Z.~Huang, P.~Mathur, J.~Yuan, D.~Bekal, X.~Niu, S.~M. Jayanthi \emph{et~al.}, ``Speechverse: A large-scale generalizable audio language model,'' \emph{arXiv preprint arXiv:2405.08295}, 2024.

\bibitem{rubenstein2023audiopalm}
P.~K. Rubenstein, C.~Asawaroengchai, D.~D. Nguyen, A.~Bapna, Z.~Borsos, F.~d.~C. Quitry, P.~Chen, D.~E. Badawy, W.~Han, E.~Kharitonov \emph{et~al.}, ``Audiopalm: A large language model that can speak and listen,'' \emph{arXiv preprint arXiv:2306.12925}, 2023.

\bibitem{du2023lauragpt}
Z.~Du, J.~Wang, Q.~Chen, Y.~Chu, Z.~Gao, Z.~Li, K.~Hu, X.~Zhou, J.~Xu, Z.~Ma \emph{et~al.}, ``Lauragpt: Listen, attend, understand, and regenerate audio with gpt,'' \emph{arXiv preprint arXiv:2310.04673}, 2023.

\bibitem{zhang2023speechgpt}
D.~Zhang, S.~Li, X.~Zhang, J.~Zhan, P.~Wang, Y.~Zhou, and X.~Qiu, ``Speechgpt: Empowering large language models with intrinsic cross-modal conversational abilities,'' \emph{arXiv preprint arXiv:2305.11000}, 2023.

\bibitem{nguyen2025spirit}
T.~A. Nguyen, B.~Muller, B.~Yu, M.~R. Costa-Jussa, M.~Elbayad, S.~Popuri, C.~Ropers, P.-A. Duquenne, R.~Algayres, R.~Mavlyutov \emph{et~al.}, ``Spirit-lm: Interleaved spoken and written language model,'' \emph{Transactions of the Association for Computational Linguistics}, vol.~13, pp. 30--52, 2025.

\bibitem{bapna2022mslam}
A.~Bapna, C.~Cherry, Y.~Zhang, Y.~Jia, M.~Johnson, Y.~Cheng, S.~Khanuja, J.~Riesa, and A.~Conneau, ``mslam: Massively multilingual joint pre-training for speech and text,'' \emph{arXiv preprint arXiv:2202.01374}, 2022.

\bibitem{chu2024qwen2}
Y.~Chu, J.~Xu, Q.~Yang, H.~Wei, X.~Wei, Z.~Guo, Y.~Leng, Y.~Lv, J.~He, J.~Lin \emph{et~al.}, ``Qwen2-audio technical report,'' \emph{arXiv preprint arXiv:2407.10759}, 2024.

\bibitem{michel2019sixteen}
P.~Michel, O.~Levy, and G.~Neubig, ``Are sixteen heads really better than one?'' \emph{Advances in neural information processing systems}, vol.~32, 2019.

\bibitem{simonyan2013deep}
K.~Simonyan, A.~Vedaldi, and A.~Zisserman, ``Deep inside convolutional networks: Visualising image classification models and saliency maps,'' \emph{arXiv preprint arXiv:1312.6034}, 2013.

\bibitem{prabhavalkar2018minimum}
R.~Prabhavalkar, T.~N. Sainath, Y.~Wu, P.~Nguyen, Z.~Chen, C.-C. Chiu, and A.~Kannan, ``Minimum word error rate training for attention-based sequence-to-sequence models,'' in \emph{2018 IEEE International Conference on Acoustics, Speech and Signal Processing (ICASSP)}.\hskip 1em plus 0.5em minus 0.4em\relax IEEE, 2018, pp. 4839--4843.

\bibitem{vincent2016chime4}
E.~Vincent, S.~Watanabe, J.~Barker, and R.~Marxer, ``The 4th chime speech separation and recognition challenge,'' \emph{URL: http://spandh. dcs. shef. ac. uk/chime\_challenge/(last accessed on 1 August, 2018)}, 2016.

\bibitem{font2013freesound}
F.~Font, G.~Roma, and X.~Serra, ``Freesound technical demo,'' in \emph{Proc. ACM MM}, 2013, pp. 411--412.

\bibitem{ardila2019common}
R.~Ardila, M.~Branson, K.~Davis, M.~Henretty, M.~Kohler, J.~Meyer, R.~Morais, L.~Saunders, F.~M. Tyers, and G.~Weber, ``Common voice: A massively-multilingual speech corpus,'' \emph{arXiv preprint arXiv:1912.06670}, 2019.

\bibitem{2020swb}
Z.~Tüske, G.~Saon, K.~Audhkhasi, and et~al, ``Single headed attention based sequence-to-sequence model for state-of-the-art results on switchboard,'' in \emph{arXiv preprint arXiv:2001.07263}, 2020.

\bibitem{wang2020covost}
C.~Wang, J.~Pino, A.~Wu, and J.~Gu, ``Covost: A diverse multilingual speech-to-text translation corpus,'' \emph{arXiv preprint arXiv:2002.01320}, 2020.

\bibitem{zhao2024librisqa}
Z.~Zhao, Y.~Jiang, H.~Liu, Y.~Wang, and Y.~Wang, ``Librisqa: A novel dataset and framework for spoken question answering with large language models,'' \emph{IEEE Transactions on Artificial Intelligence}, 2024.

\bibitem{rafailov2024direct}
R.~Rafailov, A.~Sharma, E.~Mitchell, C.~D. Manning, S.~Ermon, and C.~Finn, ``Direct preference optimization: Your language model is secretly a reward model,'' \emph{Advances in Neural Information Processing Systems}, vol.~36, 2024.

\end{thebibliography}

\end{document}